# Modeling and Analysis of Loading Effect in Leakage of Nano-Scaled Bulk-CMOS Logic Circuits*

Saibal Mukhopadhyay, Swarup Bhunia, and Kaushik Roy

Dept. of ECE, Purdue University, West Lafayette, IN-47907, USA

<sm, bhunias, kaushik>@ecn.purdue.edu

## ABSTRACT

*In nanometer scaled CMOS devices significant increase in the subthreshold, the gate and the reverse biased junction band-to-band-tunneling (BTBT) leakage, results in the large increase of total leakage power in a logic circuit. Leakage components interact with each other in device level (through device geometry, doping profile) and also in the circuit level (through node voltages). Due to the circuit level interaction of the different leakage components, the leakage of a logic gate strongly depends on the circuit topology i.e. number and nature of the other logic gates connected to its input and output. In this paper, for the first time, we have analyzed loading effect on leakage and proposed a method to accurately estimate the total leakage in a logic circuit, from its logic level description considering the impact of loading and transistor stacking.*

## 1. INTRODUCTION

Aggressive scaling of CMOS devices in each technology generation has resulted in significant increase in the leakage current in CMOS devices. In nano-scaled devices the three major leakage components can be identified as: Subthreshold leakage, Gate leakage and reverse biased drain-substrate and source-substrate junction Band-To-Band-Tunneling (BTBT) leakage [1-3]. In a transistor, the relative magnitudes of these components depend on the device geometry (namely, channel length, oxide thickness and transistor width), the doping profiles and the operating temperature. In a CMOS device the different leakage components interact strongly with each other. On the other hand, the different leakage components depend on the terminal voltages of a transistor. Hence, in logic circuits leakage components interact with each other through the node voltages. If the output node (say, $N_0$) of a gate (say, G) is connected to the input of the other gates ($G_{out1}$, $G_{out2}$, ..,$G_{outn}$), the gate leakage from these other gates change the voltage at OUT1 (Fig. 1). This effect can be defined as the "loading effect" and it modifies the leakage of the gate G, $G_{out1}$, …, $G_{outn}$.

In this work, we have analyzed the impact of "loading effect" on the leakage of a logic gate and logic circuit. In particular, in this paper:

- We have described the interaction of different leakage components in a device and in a circuit.
- We have evaluated effect of loading on the individual leakage components and the total leakage of a logic gate.
- We have proposed a methodology to efficiently and accurately estimate the total leakage of a logic circuit from its gate-level description considering loading effect.

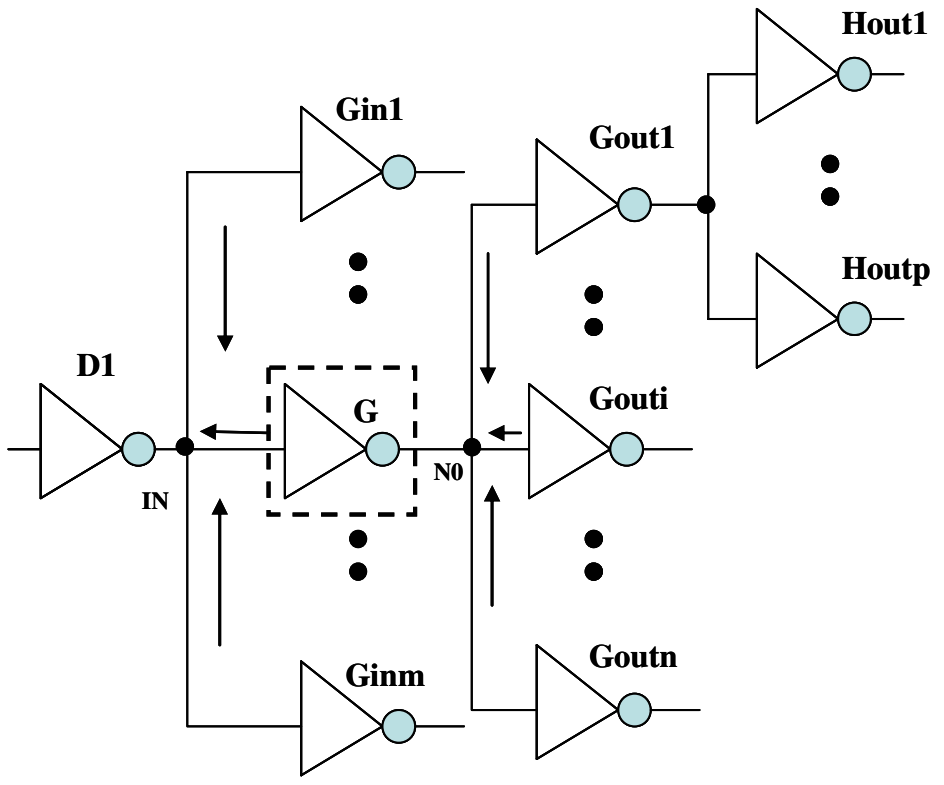

**Fig 1. Illustration of loading effect for the logic gate G.**

*This work is supported in part by Semiconductor Research Corporation, MARCO GSRC, Intel, and IBM Corp.

## 2. LEAKAGE COMPONENTS IN A DEVICE

### 2.1 Device Structures

Leakage analysis presented in this work is based on the transistors of 50nm gate length designed using the device simulator MEDICI [4]. The device structure and "super halo" doping profiles given in [5] were used in designing the transistors. The parameter extraction tool, AURORA [6] was used to extract BSIM4 SPICE model parameters of the designed devices to do SPICE simulations.

### 2.2 Leakage Components

In nano-scaled CMOS devices the major leakage components are shown in Fig. 2. The details of individual leakage components are given below:

**(1) Subthreshold current ($I_{ds}$):**

The subthreshold current in a transistor is caused by the diffusion of the minority carriers from the source to the drain. The subthreshold current depends exponentially on the threshold voltage of a transistor. In nano-scaled devices the short channel effects (penetration of the drain electric field into the channel) (SCE) reduces the threshold voltage thereby increasing the subthrshold current [1-3]. Due to SCE, the subthreshold current increases with an increase in the drain bias (Drain Induced Barrier Lowering) and reduction in the channel length (Vth-roll off). At a high oxide electric field (sub-100nm regime), the quantization of the electron energy in the channel region tends

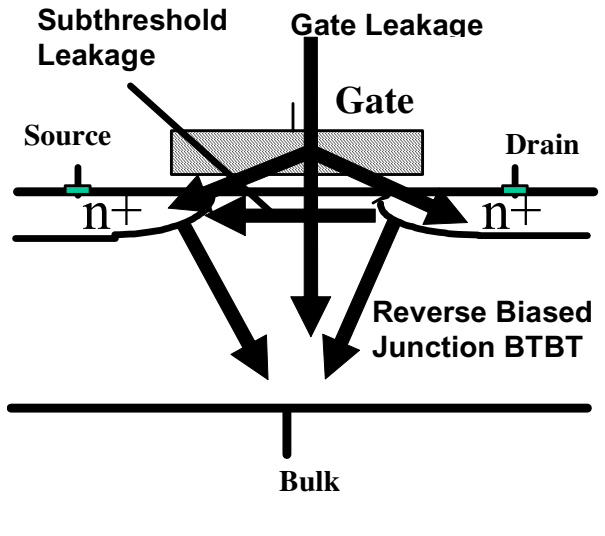

**Fig.2 Leakage components in a device**

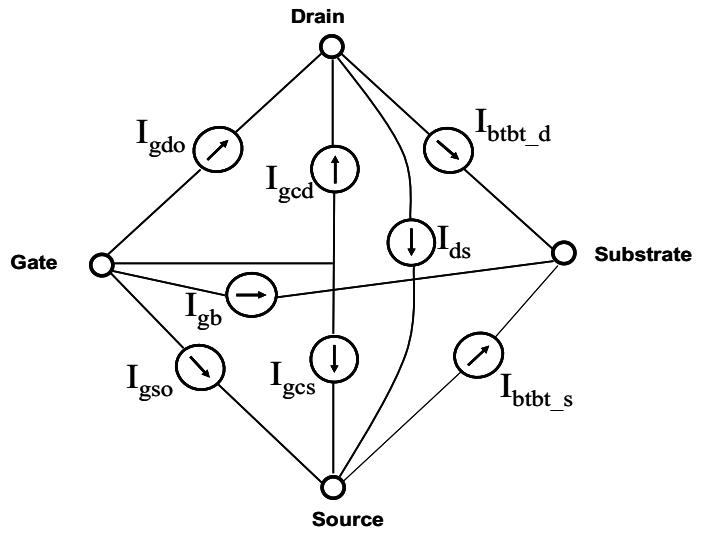

**Fig.3. Leakage current sources in a transistor**

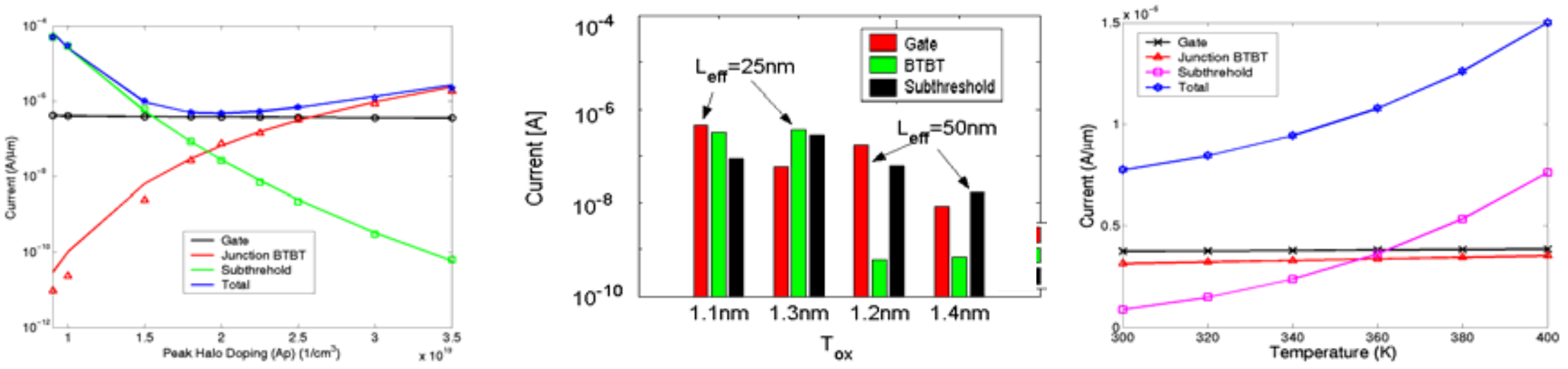

**(a)** **(b)** **(c)**

**Fig. 4: Variation of different leakage current components with the (a) Halo doping and (b) device geometry, and (c) temperature.**

to increase the threshold voltage, thereby lowering the subthreshold current [1-3].

**(2) Gate Direct Tunneling Current ($I_{gate}$)**

At ultra-thin gate oxide regime, due to the high electric field and low oxide thickness, electrons can tunnel through the gate oxide. This is known as the direct tunneling of electrons and results in a large gate leakage in nano-scale transistors. An increase in the supply voltage and/or reduction in the oxide thickness, result in an exponential increase in the gate tunneling current [1-3,7]. Major components of gate tunneling in a scaled MOSFET are: (a) gate to S/D overlap region current components ($I_{gso}$ & $I_{gdo}$), (b) gate to channel current ($I_{gc} = I_{gcs} + I_{gcd}$), and (c) gate to substrate current ($I_{gb}$) [7].

**(3) Junction Band-To-Band-Tunneling current ($I_{JN}$)**

Application of a reverse bias across the highly doped p-n junction results in the tunneling of electrons from the valence band of p-side to the conduction band of n-side [1-3]. This is known as junction band-to-band-tunneling (BTBT) current. In nano-scale MOSFETs due to the use of high junction doping ("Halo" implants used to suppress SCE), large junction BTBT occurs at "off" state with drain at $V_{DD}$ and substrate at ground (at high drain-to-substrate reverse bias) [1-3]. The junction BTBT exponentially increases with an increase in the junction doping and supply voltage. [1-3]

### 2.3. Total Leakage in a Transistor

The total leakage in a device is the summation of the three major leakage components ($I_{total}=I_{BTBT} + I_{sub} + I_{gate}$). For leakage estimation the device can be modeled as a combination of voltage controlled current sources where each current source represents a current component (Fig. 3) [2].

## 3. INTERDEPENDENCE OF LEAKAGE COMPONENTS IN A DEVICE

The subthreshold, the gate and the junction BTBT leakage depend on each other through device geometry (particularly oxide thickness), doping profile and temperature. Increasing the Halo doping concentration increases the junction BTBT (by increasing the junction field) whereas it reduces the subthreshold current (by reducing the short channel effect) [1-3] (Fig. 4(a)). The gate leakage is insensitive to halo doping concentration. Increasing the oxide thickness reduces the gate leakage. Higher oxide thickness also increases the short channel effect, thereby increasing the subthreshold leakage in nano-scaled transistors [1-3] (Fig. 4(b)). The junction BTBT is not a strong function of the oxide thickness.

The different leakage components show different temperature dependence. Subthreshold current increases exponentially with temperature whereas the gate tunneling current is almost insensitive to temperature variation. Due to the reduction of the band-gap at a higher temperature, junction BTBT increases (marginally) with temperature (Fig. 4(c)) [1-3]. It can be observed that, at room temperature (T=300K) the gate leakage and the junction BTBT dominate over subthreshold current, while at an elevated temperature, subthreshold leakage is the dominant component of the overall leakage. Hence, in active mode, subthreshold is the major component of leakage.

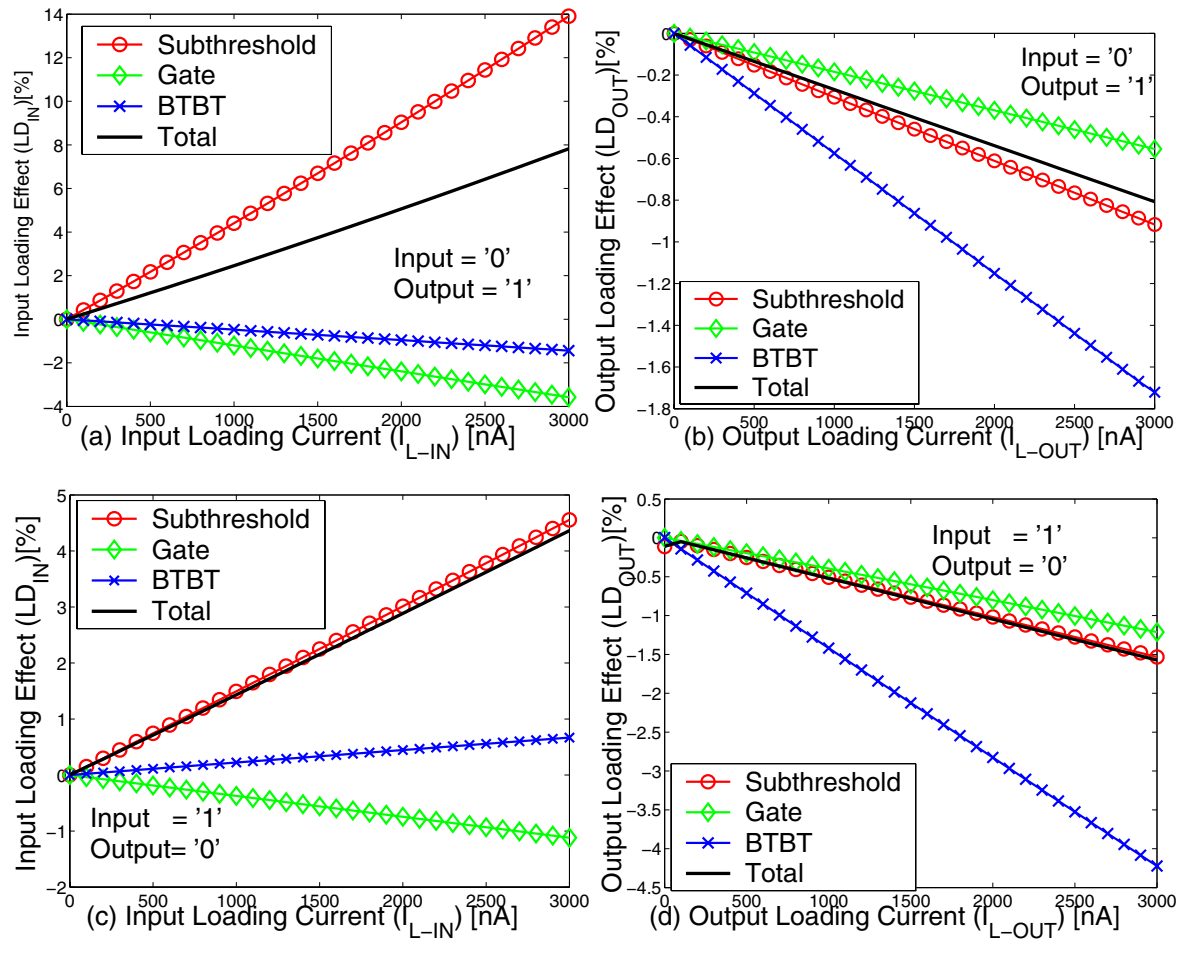

**Fig. 5: Input and output loading effect with different inputs to an inverter.**

## 4. INTERACTION OF DIFFERNENT LEAKAGE COMPONENTS IN CIRCUITS

In a logic circuit different leakage components interact with each other through the internal nodes. Such an interaction changes the internal node voltages in a circuit and hence, modifies the leakage of individual logic gates and a logic circuit. The interaction of different leakage components determines the leakage of a logic gate due to "stacking effect" [8], [9]. It has been shown in [8] that the leakage of a logic gate at different input vectors depends on the relative strength of the different leakage components in a device. For example, for a subthreshold leakage dominated device, the minimum leakage vector in a 2-input NAND gate is "00", while, for a gate leakage dominated device it is "10". In [2], authors have discussed the method to estimate the leakage of a logic gate considering the interaction of leakage components within a logic gate (*intra-gate* interaction). However, it does not consider the interaction of leakage components of different logic gates (*inter-gate* interaction).

To understand the impact of inter-gate interaction of leakage at the circuit level, let us consider the circuit shown in Fig. 1. The leakage of inverter G can be evaluated by solving Kirchhoff's

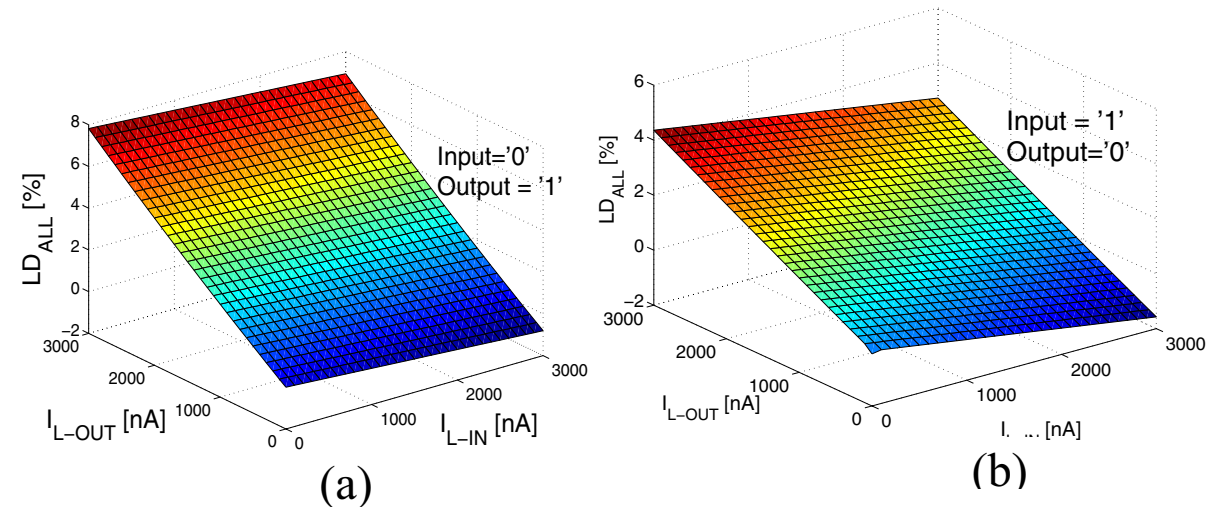

**Fig. 6: Variation of leakage due to loading at both input and output with: (a) input '0' and (b) input '1'.**

Current Law (KCL) at the node $N_0$ (i.e. output node of G), given by (Fig. 1).

$$I_{ddP} + I_{ddN} = 0$$
$$\Rightarrow (I_{dsN} + I_{gcdN} + I_{gdoN} - I_{jnN}) + (I_{dsN} + I_{gcdN} + I_{gdoN} - I_{jnN}) = 0 \quad (1)$$

Voltage at $N_0$ ($V_{N0} = f(I_{ddP}, I_{ddN})$) can be obtained by solving (1). However, since node $N_0$ is also connected to the inputs of inverters $G_1$ to $G_n$, the gate leakage from these inverters will be added to the node $N_0$. Hence, the KCL at node OUT will modify to:

$$I_{ddP} + I_{ddN} + \sum_{i = load\ gates} \left[ I_{gateN-Gi} + I_{gateP-Gi} \right] = 0 \quad (2)$$

Hence, $V_{N0}$ also depends on the gate leakage of the NMOS ($I_{gateN-Gi}$) and PMOS ($I_{gateP-Gi}$) of the inverters $G_1$ to $G_N$ ($V_{N0}=f(I_{ddP}, I_{ddN}, I_{gateN-Gi}, I_{gateN-Gi})$).

If $\sum (I_{gateN-Gi} + I_{gateP-Gi}) \neq 0$, $V_{N0}$ obtained from (1) will be different from the $V_{N0}$ obtained from (2). The modification of $V_{N0}$ has following effects:

- A change in the output voltage ($V_{OUT\text{-}G} \equiv V_{N0}$) of inverter G will modify its gate, subthreshold and junction BTBT leakage.
- A change in the input voltage ($V_{IN\text{-}Gi} \equiv V_{N0}$) of inverters $G_1$ to $G_N$ will also modify their subthreshold (principally) and gate leakage.

Hence, the leakage of a gate depends on its *input loading* (i.e. total gate leakage of other gates connected to its input node $I_{L\text{-}IN}$) and *output loading* (total gate leakage of other gates connected to its output node $I_{L\text{-}OUT}$). Thus we can define the *input* ($LD_{IN}$) and *output* ($LD_{OUT}$) *loading effect* as the change in the leakage of a logic gate due to its input and output loading, respectively. $LD_{IN}$ and $LD_{OUT}$ can be expressed as:

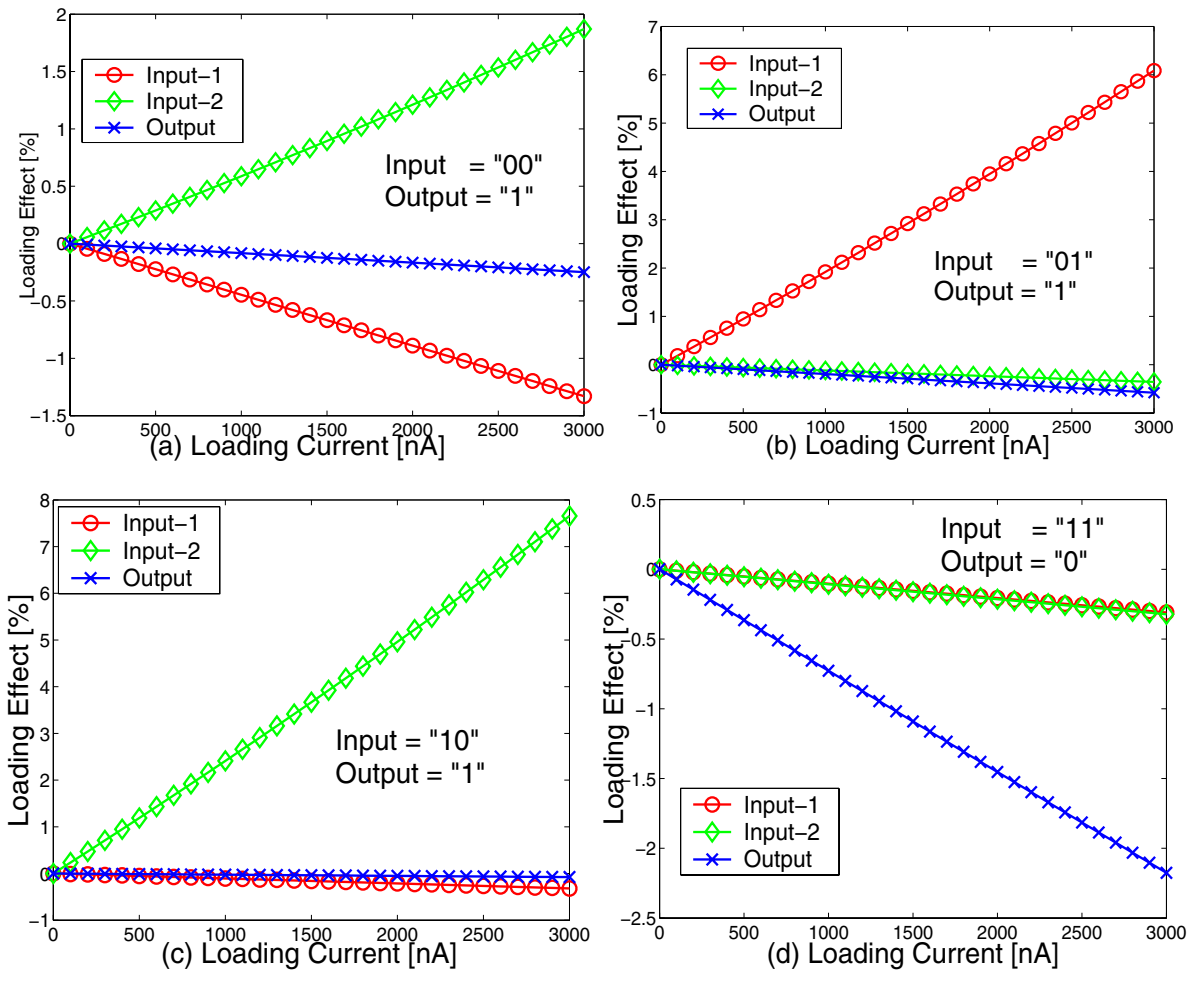

**Fig. 7: Loading effect (input and output) in the total leakage of a NAND gate under different input vectors.**

$$LD_{IN}(I_{L-IN}) = \frac{L_G(I_{L-IN}) - L_{NOM}}{L_{NOM}}$$
$$LD_{OUT}(I_{L-OUT}) = \frac{L_G(I_{L-OUT}) - L_{NOM}}{L_{NOM}} \quad (3)$$

where, $L_{NOM}$ is the nominal leakage of a gate in isolation (i.e. without any output and input loading); $L_G(I_{L\text{-}IN})$ is the leakage of the gate with input loading $I_{L\text{-}IN}$; and $L_G(I_{L\text{-}OUT})$ is the leakage of the gate with output loading $I_{L\text{-}OUT}$. The overall loading effect ($LD_{ALL}$) depends on both input and output loading and given by:

$$LD_{ALL}(I_{L-IN}, I_{L-OUT}) = \frac{L_G(I_{L-IN}, I_{L-OUT}) - L_{NOM}}{L_{NOM}} \quad (4)$$

For the logic gates with multiple inputs, there will be a $LD_{IN}$ associated with each input. Hence, the $LD_{ALL}$ for a multiple input gate is given by:

$$LD_{ALL}(I_{L-IN-k}, I_{L-OUT}) = \frac{L_G(I_{L-IN-k}, I_{L-OUT}) - L_{NOM}}{L_{NOM}} \quad (5)$$
$$\forall k = 1, 2, \ldots, N, where,\ N\ is\ the\ \#of\ inputs$$

Fig. 5 shows the variation of $LD_{IN}$, $LD_{OUT}$ and $LD_{ALL}$ of an inverter with a variation in the $I_{L\text{-}IN}$ and $I_{L\text{-}OUT}$ at different input conditions for the 25nm device. It can be observed that, the impact of loading increases with increase in the loading currents. Moreover, *LD* values strongly depend on the input condition of the inverter. The total leakage of the inverter with input '0' and '1' is given by:

$$Input = 0, Output = 1:$$
$$I_{INV_0} = I_{sub_N} + I_{JN_N} + I_{gdo_N} + I_{g_P} (= I_{gc_P} + I_{gdo_P} + I_{gso_P}) \quad (6)$$
$$Input = 1, Output = 0:$$
$$I_{INV_1} = I_{sub_P} + I_{JN_P} + I_{gdo_P} + I_{g_N} (= I_{gc_N} + I_{gdo_N} + I_{gso_N})$$

Due to input loading, the voltage at the input node of the inverter gets modified. If input of G in Fig. 1 is at '0' (i.e. $V_{IN}$=0V), the gate currents of input loading gates $G_{IN1},\ldots,G_{Inn}$ increases $V_{IN}$ from 0V. If the input is at '1' (i.e. $V_{IN}=V_{DD}$) the gate leakage through $G_{IN1},\ldots, G_{INn}$ reduces the voltage from $V_{DD}$. This increases the $|V_{GS}|$ of the 'off' transistor (i.e. NMOS at input='0' and PMOS at input='1') thereby increasing the subthreshold leakage of the inverter G. On the other hand, it marginally reduces the gate currents of the PMOS and NMOS by reducing the $|V_{GD}|$ (PMOS and NMOS) and $|V_{GS}|$ (of PMOS at input='0' and NMOS at input='1'). Since the junction leakage is a weak function of the gate voltage, input loading has minimal impact on the junction leakage [2]. Similarly, for output loading, $V_{OUT}$ increases from 0V when output is '0' and decreases from $V_{DD}$ when output is '1'. This reduces (a) the $|V_{DS}|$ of the 'off' transistor (PMOS when output= '0' and NMOS when output='1'), thereby reducing the subthreshold leakage; (b) $|V_{GD}|$ of the PMOS and NMOS thereby reducing the gate leakage; and (c) $|V_{DB}|$ of the transistor contributing to the junction BTBT (e.g. PMOS when output is '0' and NMOS when output is '1'), thereby reducing the junction leakage. Hence, due to input loading subthreshold leakage increases while gate leakage reduces and junction leakage remains almost constant (Fig. 5). On the other hand, due to output loading all the three components of the leakage reduces (Fig. 5).

It can also be observed from Fig. 5 that, the input loading effect ($LD_{IN}$) is most pronounced in the subthreshold leakage as it changes the $V_{gs}$ of the 'off' transistor. Output loading ($LD_{OUT}$) has the strongest impact on the junction leakage by changing $|V_{DB}|$ of the transistor which contributes to the junction BTBT. The gate leakage experience minimum change due to the loading effect.

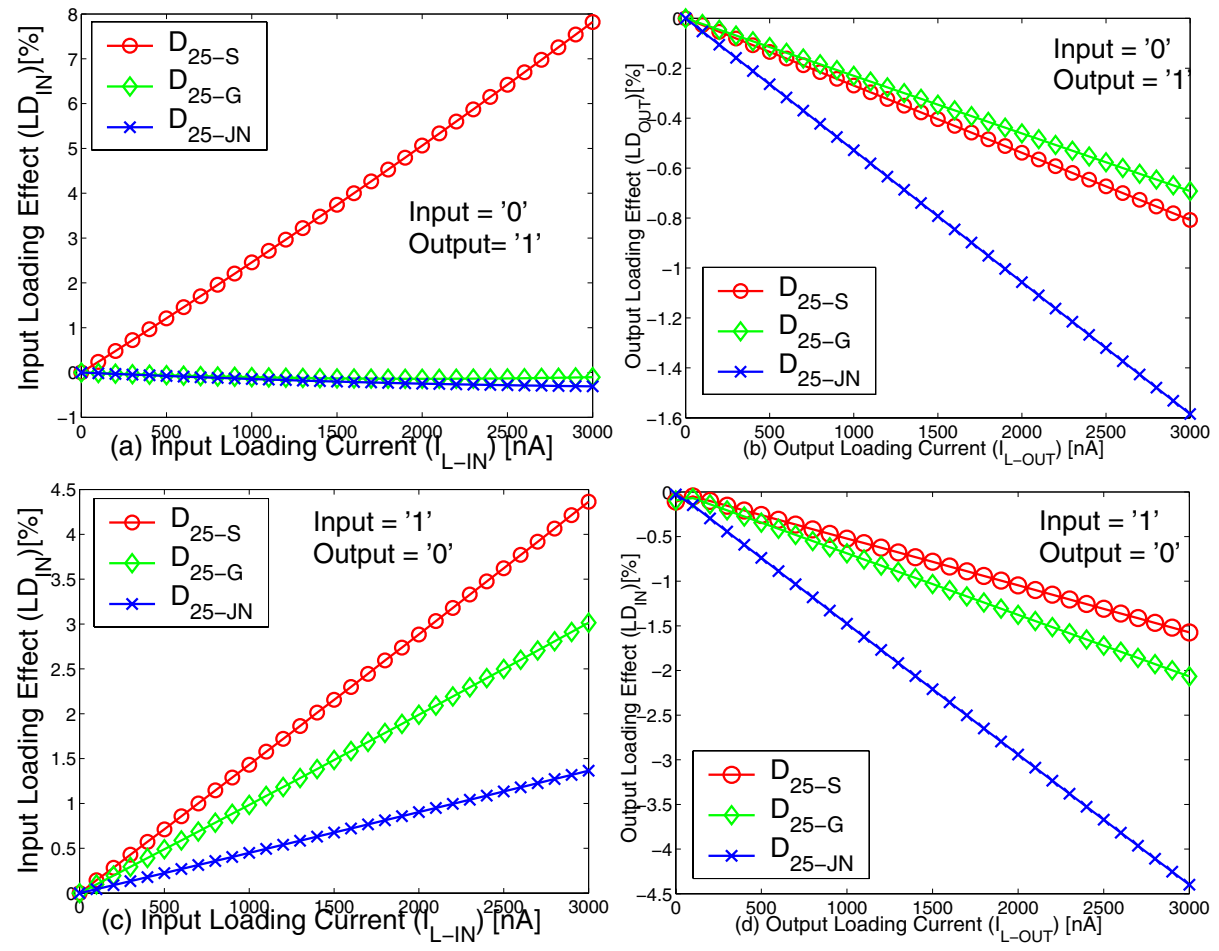

**Fig. 8: Loading effect for different devices: (a) input & (b) output loading effect with input '0'; and (c) input & (d) output loading effect with input '1'.**

Input loading effect (both on the total leakage and on the subthreshold leakage) is more observable with input '0' (Fig. 5). This is due to the fact that, with input '0' the subthreshold leakage is principally contributed by the 'off' NMOS transistor, whereas 'off' PMOS transistor determines the subthreshold leakage with input '1'. Since short channel effect is more serious in PMOS [6], the subthreshold leakage in PMOS is less sensitive to $V_{gs}$ than NMOS. $V_{ds}$ sensitivity of PMOS subthreshold leakage is higher than that of NMOS subthreshold current. Since the output loading modifies the $V_{ds}$ of a transistor it has a stronger impact on PMOS (i.e. when input='1' and output='0') subthreshold leakage. The impact of output loading on junction BTBT is also stronger with output '0' (PMOS contributes to junction BTBT) than with output '1' (NMOS determines junction BTBT). This is due to the fact that, PMOS has a larger junction BTBT current [2]. Consequently, effect of output loading is higher with output at '0'. Fig. 6 shows the loading effect considering both the input and output loading (i.e. $LD_{ALL}$). It can be observed that $LD_{ALL}$ is normally higher with input = '0'.

Analysis of loading effect on the 2-input NAND gate shows the input vector dependence of the loading effect (Fig. 7). From Fig. 7 it can be observed that, input loading is higher if at least one of the inputs is at '0' (i.e. with vectors '00', '01', '10'). This is because of the fact that, the input loading has a stronger effect on the subthreshold leakage of an 'off' NMOS. Due to the reduction in the subthreshold leakage by the stacking effect [9], input loading has less effect with input '00' compared to the inputs '01' or '10'. As observed in the case of the inverter, the effect of output loading is higher with the output equal to '0'. Moreover, depending on the input vector the loading effect may increase or reduce the total leakage of the gate.

From the above discussion it can be concluded that:

- The loading of a logic gate changes the different leakage components and the total leakage.
- The gate leakage is the cause of the loading effect. However, its effect is mostly observed in the subthreshold and the junction BTBT leakage.
- The loading effect depends on the input and output logic levels, the magnitude, and the relative strength of the different leakage components in PMOS and NMOS.

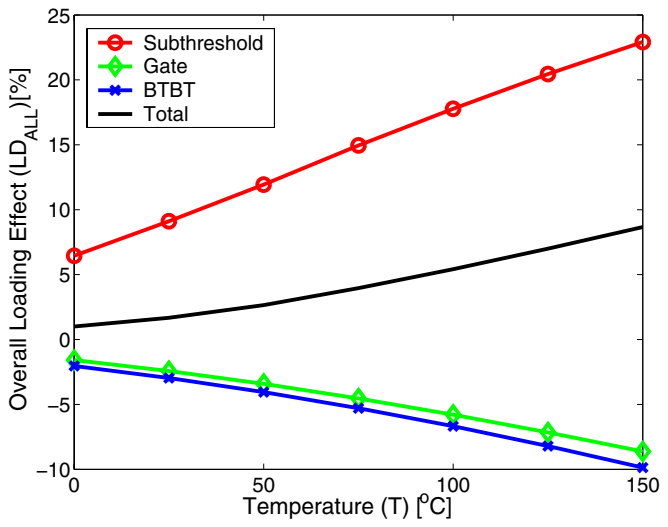

**Fig. 9: Impact of temperature on the overall loading effect (*$LD_{ALL}$*) of an Inverter (input= '0', output='1').**

## 5. VARIATION IN THE LOADING EFFECT

In this section, we discuss the impact of change in relative strengths of different leakage components with device design, temperature and process parameter variations on loading effect.

### 5.1. Effect of relative strengths of leakage components

Fig. 8 shows the input and output loading effect of an inverter designed with devices with different relative strengths of the leakage components. The subthreshold leakage, the gate leakage and the junction leakage dominates the total leakage in device $D_{25\text{-}S}$, $D_{25\text{-}G}$ and $D_{25\text{-}JN}$, respectively (total leakage is same in the three devices). It has been discussed earlier that, the input loading has the strongest impact on the subthreshold leakage. Hence, the input loading effect is most pronounced in the inverter designed with $D_{25\text{-}S}$ (subthreshold leakage dominated device). Input loading has a weaker impact on the inverters designed with $D_{25\text{-}JN}$ (junction leakage dominated device) and $D_{25\text{-}G}$ (gate leakage dominated device). On the other hand, output loading effect is most pronounced in the inverter designed with $D_{25\text{-}JN}$ (since, junction BTBT is the strongest function of the output loading among the three different leakage components). In general the loading has least impact on the gate leakage dominated device.

### 5.2. Impact of temperature on the loading effect.

Since the gate leakage is a weak function of temperature (Fig. 4c), it can be concluded that the "*cause of the loading*" does not increase significantly with temperature. But, the "*effect of loading*" (i.e. the subthreshold and the junction tunneling current) are strong function of temperature. With the increase in the temperature the effect of loading on the subthreshold leakage significantly increases (Fig. 9). An increase in temperature exponentially increases the subthreshold leakage in device. This has a two fold impact on the loading effect. First, the increase in the subthreshold leakage due to an increase in the $|V_{GS}|$ of the NMOS in the inverter G is higher at a higher temperature. Second, the contribution of the subthreshold current and the junction current of the PMOS of the inverter D to node IN (i.e. input of G and output of D), increases at a higher temperature. Hence, the voltage- rise in the node IN (i.e. $|V_{GS}|$ of NMOS) increases, thereby increasing the subthreshold current considerably. However, the increase in the input voltage of the inverter G reduces its output voltage (due to larger subthreshold current of NMOS). This reduction in the output voltage coupled with the increase in the input voltage, reduces the gate and the junction BTBT current (as explained in section. 4). Thus, the loading effect for the gate and the junction BTBT also increases with the temperature. However, since the subthreshold, the gate, and the junction BTBT moves in the reverse direction with the increase in the temperature, the impact of the temperature on the loading-effect of the total leakage is less significant.

### 5.3. Impact of process variation

The variation in the process parameters (e.g. channel length ($L$), oxide thickness ($T_{ox}$), threshold voltage ($V_{th}$), supply voltage ($V_{DD}$) etc.) result in a large variation in the leakage in transistors and logic gates [10]. The subthreshold leakage is extremely sensitive to process variation, whereas, the gate leakage and the junction BTBT leakage are less sensitive [10]. The application of the random variation in $L$, $V_{th}$, and $T_{ox}$ of different transistors and in $V_{DD}$ results in the significant variation in the different leakage components and the total leakage (Fig. 10) (obtained through 10,000 Monte-Carlo simulations in SPICE). It can be observed that, the loading effect considerably modifies the leakage distributions. The maximum modification can be observed in the subthreshold leakage. It can be observed that with an increase in the inter-die variation loading effect on the mean and the standard deviation increases (Fig. 11). Particularly, consideration of the loading significantly increases the standard deviation (with σVt=50mV, loading increases the standard deviation by more than 40%). This indicates that the maximum value of leakage significantly increases due to the loading effect (almost by 2X) considering parameter variation.

The observations from the previous discussion can be summarized as follows:

- Increase in the subthreshold leakage (due to device design or increase in temperature or parameter variations) has a strong impact on the overall loading effect.
- Consideration of the loading effect significantly increases the leakage spread of a circuit under parameter variation.

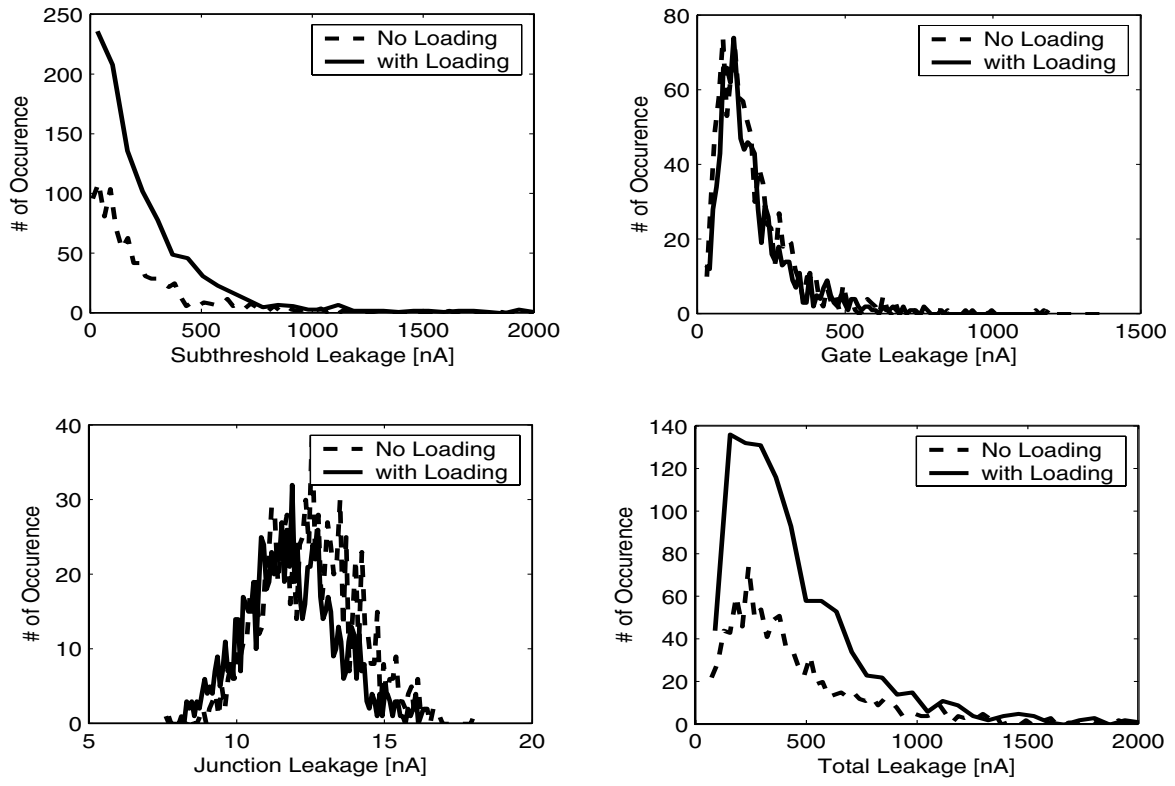

**Fig. 10: Distribution of different leakage components of an inverter (input='0' and output='1') with and without loading (input loading of 6 inverters and output loading of 6 inverters).**

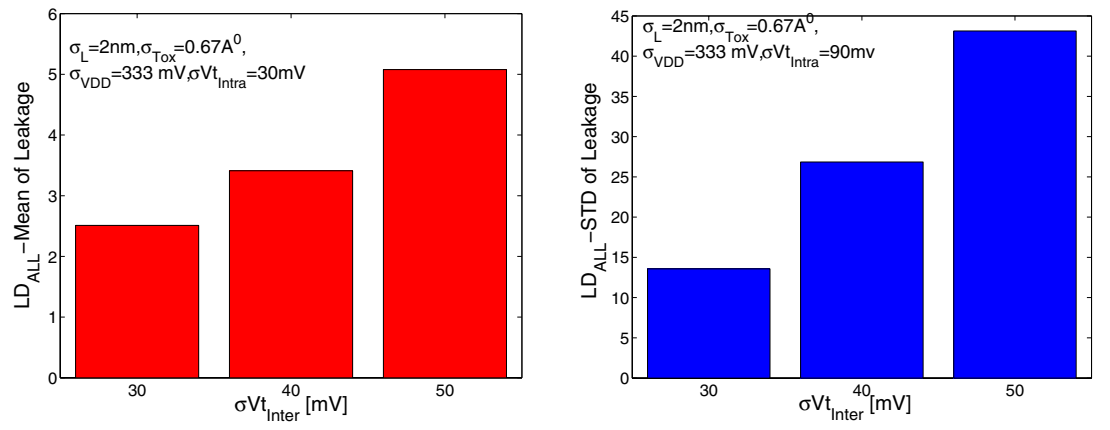

**Fig. 11: Effect of loading on the mean (left) and standard deviation (right) of the total leakage in an inverter (input='0').**

## 6. LOADING EFFECT AT CIRCUIT LEVEL

Traditionally, leakage current in a circuit is calculated by determining individual leakage values for each gate and accumulating them. This procedure is valid assuming that leakage current in a gate is independent of the circuit topology i.e. it is not affected by the leakage in other gates. However, due to loading effect, the leakage of gate depends on the leakage of the other gates. Hence, we need to consider propagation of loading effect across logic gates for accurate circuit-level estimation of leakage current in nano-scale CMOS. This is similar to the propagation of slope changes for delay calculation,

Leakage of a logic gate is related to the voltage difference in its input and output nodes due to the loading effect. The cause of the loading effect is the gate leakage and its effect is observed in the subthreshold, the junction BTBT, and the gate leakage. Let us look into the case of output loading effect for the simple circuit shown in Fig. 1 (i.e. a driver D1 that drives gates Gin1, $G_{in1}$,…, $G_{inm}$ and fanout gates $G_{out1}$, $G_{out2}$,…, $G_{outn}$). Also consider that, $G_{out1}$ is connected to the input of the inverters $H_{out1}$, $H_{out2}$,..., Houtp. For the inverter G, output voltage difference due to loading effect is determined by the leakage of the fanout gates $G_{out1}$,…,$G_{outn}$. However, leakage of the gates $G_{out1}$…, $G_{outn}$ is again determined by the leakage current of $H_{out1}$,…,$H_{outp}$. Leakage current of $H_{out1}$,..,$H_{outp}$ are, again, related to the leakage of following gates. Similar relation holds through input loading. Hence, in order to estimate circuit leakage, we need to simultaneously solve a set of KCL equations with $n$ variables, where $n$ is the number of internal nodes of the circuit.

We can, however, avoid the need for solving simultaneous equations and come up with a simple method for circuit-level leakage estimation in presence of loading effect. It has been observed that the effect of loading on the gate leakage is not very high. The gate leakage of the gates $H_{out1}$,...,$H_{outn}$ in Fig. 1 will modify the voltage at the output of $G_{out1}$, thereby modifying its leakage. However, it can be observed from earlier figures that, the change in the output voltage of $G_{out1}$ will not modify the gate leakage of $G_{out1}$ strongly. Thus, the effect of the gate leakage of $H_{out1}$,...,$H_{outn}$ on the voltage at the output of inverter G is minimal. Thus, the modification of the leakage of G due to the gate leakage of $H_{out1}$,...,$H_{outn}$ is negligible. In other words, the propagation of the loading effect (output loading effect in this case) beyond one level is negligible. A similar argument can also be made to explain that input loading effect does not propagate strongly beyond one level.

Based on the above observation, we have developed an efficient algorithm that estimates different components of leakage considering loading effect. Leakage values generated by the algorithm closely matches results obtained from spice simulations (Fig. 12a), while being about 1000X faster than spice in run time. Flow chart for the algorithm is presented in Fig. 13. We start with a graph representing the circuit, with each vertex representing a logic gate and each edge representing a net. First, the vertices in the graph are topologically sorted [11] and the leakage values are initialized to zero. Logic values are propagated through the circuit nodes for the input pattern. Then, for each node in the graph in topological order, we compute the total input and output loading current due to the gate leakage of the corresponding gates.

The algorithm is implemented in C programming language and tested on six ISCAS89 benchmark circuits, a multiplier and an 8-bit ALU and run for 100 random vectors at T=300K. The results of the leakage estimation using the proposed algorithm

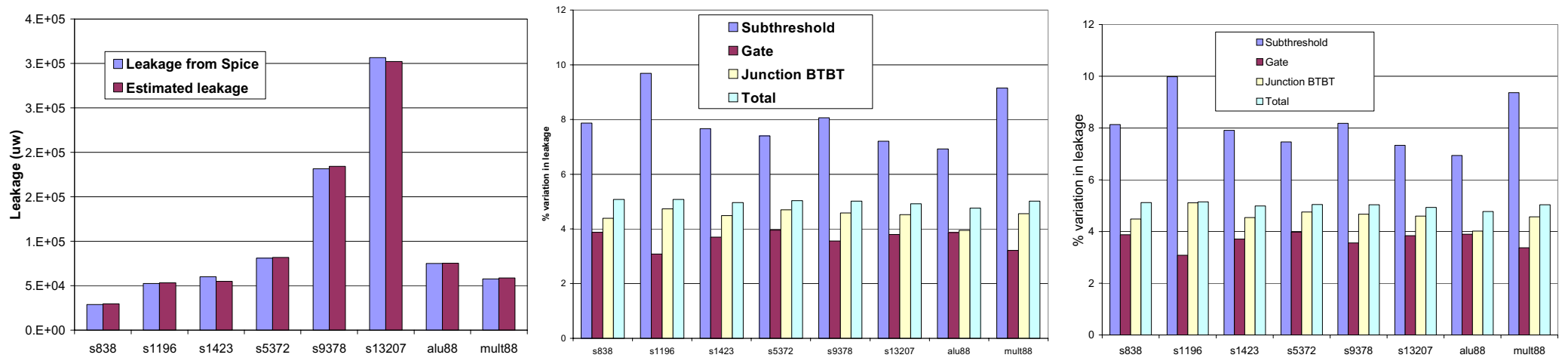

(a) (b) (c)

**Fig. 12: Estimation of leakage using the proposed procedure (a) comparison with SPICE results, (b) average leakage variation due to loading effect and (c) maximum leakage variation due to loading effect over 100 random vectors.**

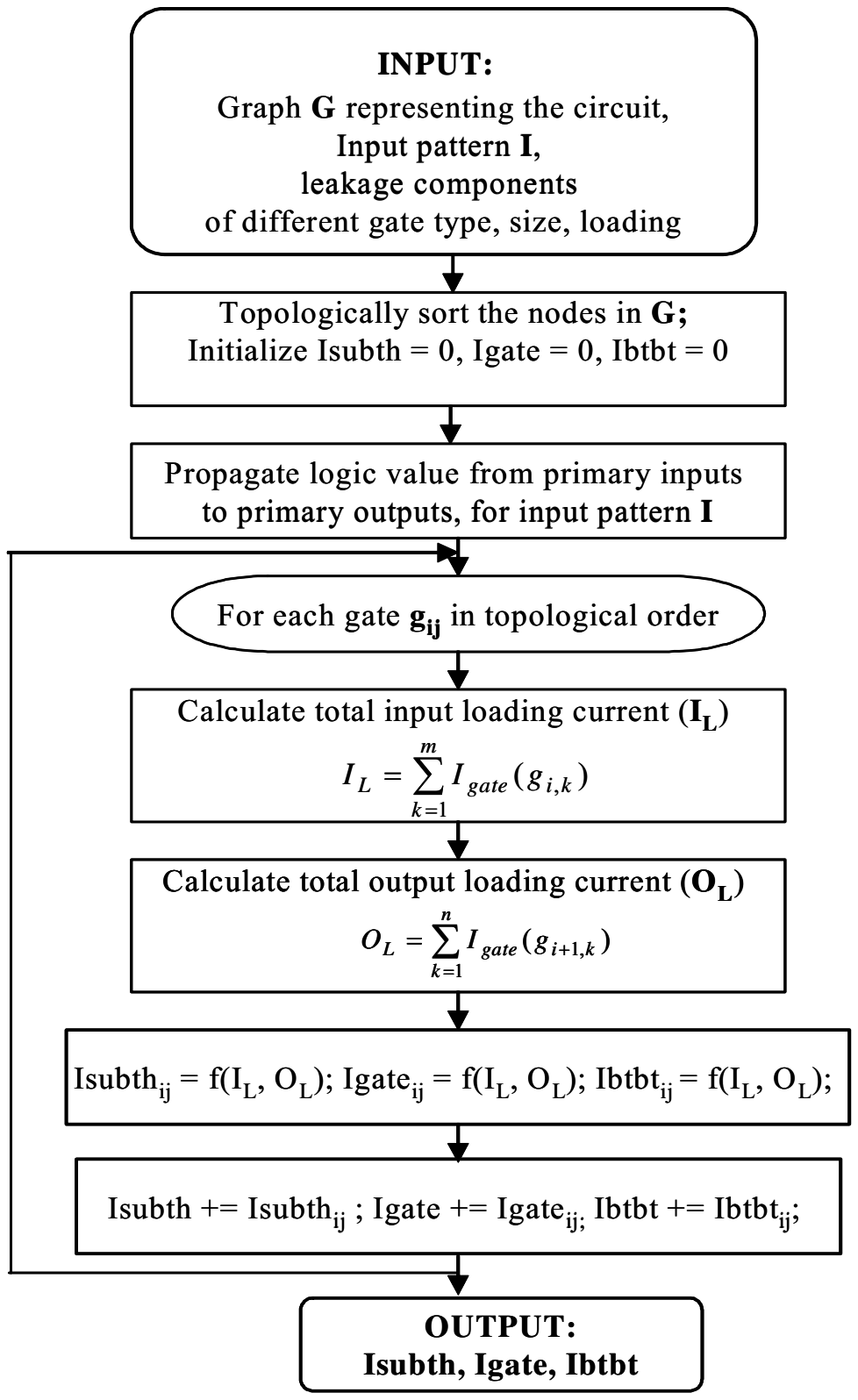

**Fig. 13: Flowchart to estimate leakage components for a circuit considering the loading effect.**

are presented in Fig. 12b and 13c. The Fig. 12b shows that the average increase in total leakage due to loading effect is about 5%. However, the variation is the sub-threshold leakage much higher (~8%) followed by the junction BTBT (~4.5%) and the gate leakage (~3.6%) (for a subthreshold leakage dominated device). Fig. 12(c) shows the respective leakage values for maximum variation due to loading over 100 vectors. The observations about the impact of loading on circuit level analysis can be summarized as follows

- Loading effect depends on circuit topology.
- The loading effect in a circuit strongly depends on the applied input pattern. The input pattern for which we obtain the minimum total leakage changes due to the loading effect. This has significant impact on the input vector control based leakage control techniques [9].

It should be noted due to the loading effect, the subthreshold leakage tends to increase while the gate and the junction BTBT tend to reduce. Moreover, in a large circuit, loading effect increases the total leakage of some logic gates while reduces that of some other gates. This is due to the input vector dependence of the loading effect (Fig. 7). Due to these factors, the overall change in the total circuit leakage due to loading is not very high (~5%).

## 7. CONCLUSIONS

In this paper, we have analyzed loading effect (caused by the voltage difference at the input/output nodes of a logic gate due to gate leakage of the gates connected to its input/output) on different leakage components of a circuit. Interaction of leakage components in a circuit through loading effect is investigated. We have demonstrated that loading effect varies with temperature and parameter fluctuations. We have presented an algorithm for fast and accurate estimation of circuit leakage considering the loading effect. Our analysis shows that, the loading effect modifies the leakage of a logic gate by 8-10%. However, in a large circuits, depending on the input vector, leakage of different logic gates moves in different directions (some increases and some reduces). In our experiments, we observed that, due to this cancellation effect, the net change in the overall leakage due to loading effect is about 5% in large circuits.